\input harvmac
\pretolerance=10000

\Title{HWS-9718, NIKHEF-96-035, hep-th/9707147}
{Anyon Statistics and the Witten Index}

\centerline{Donald Spector\footnote{$^\dagger$}
{spector@hws.edu}}
\bigskip\centerline{NIKHEF}
\centerline{Kruislaan 409, P.O. Box 41882}
\centerline{1009 DB Amsterdam, The Netherlands}
\centerline{and}
\centerline{Department of Physics, Eaton Hall}
\centerline{Hobart and William Smith Colleges}
\centerline{Geneva, NY 14456 USA\footnote{$^\ddagger$}{Permanent
address}}

\vskip .3in
Using the theory of supersymmetric anyons, I extend the definition of
the
Witten index to $2+1$ dimensions so as to accommodate the existence
of anyon spin and statistics.  I then demonstrate that, although in
general
the index receives irrational and complex contributions from anyonic
states,
the overall index is always integral, and I consider some of the
implications and interpretations of this result.

\Date{6/97; revised 11/97}

%if you want double-space, use e.g. \baselineskip=20pt plus 2pt minus
2pt

\newsec{Introduction}

The Witten index \ref\windex{E. Witten, {\it Nucl. Phys.} 
{\bf B202} (1982) 253.}
has proven 
to be a useful tool for understanding the dynamics of supersymmetric
theories, 
as well as for finding deep and illuminating connections between physics
and a 
variety of areas of mathematics. The conventional informal definition of
the 
Witten index is $\tr (-1)^F$, which can be regulated into the more
rigorously 
defined object, $\tr (-1)^F e^{-\beta H}$. Because of the degeneracy
between 
bosonic and fermionic states in supersymmetric theories, the states of
non-zero 
energy give a net zero contribution to the Witten index, and so the
Witten 
index 
is equal to the difference between the number of bosonic and the number
of 
fermionic zero energy states, which is an integer. (In a theory with a 
continuous spectrum, a fractional value is possible, due to subtleties
in the 
cancellations between densities of states \ref\akhoury{R. Akhoury and A.
Comtet,
{\it Nucl. Phys.} {\bf B246} (1984) 253.}.)

In $2+1$ dimensions, however, states of any real spin and statistics 
can arise \ref\anyons{J. M. Leinass and J.
Myrheim, {\it Nuovo Cimento} {\bf 37}, 1 (1977).}.
A standard way of obtaining exotic spin and statistics is
to couple the fields to a gauge field with Chern-Simons term.
What becomes of the index in this context?  This paper examines
this question.
We will first define the Witten index in a way that incorporates
anyons, obtaining an expression that generically involves complex
and irrational terms.  In the subsequent section, we show that,
nonetheless, regardless of the presence of Chern-Simons terms and
anyons,
 the index is an integer.  This result then motivates
the definition of an alternative index, using a grading
based on whether fields satisfy canonical commutation or anticommutation
relations.  This alternative 
index is not
as fundamental, but it is 
explicitly a sum of integers, and it necessarily has the
same value as the fundamental anyon index in any quantized field theory,
and so helps us understand the integral nature of the fundamental Witten 
index in theories with anyons. 

That the Witten index is integer-valued 
even in anyonic theories is, perhaps,
a disappointing result, as exotic values for the Witten index would be
most 
interesting; on the other hand, the fact that whatever states with
exotic spin
and statistics do arise must do so in way that will produce an integral
index
is a potentially informative constraint on the behavior of
anyonic field theories, and is worthy of careful consideration in
particular
models.  

Before continuing, the reader should note
that this paper is about the effects of Chern-Simons terms on the Witten
index, and so I do not discuss those issues which may be
relevant to the index but that are not pertinent to the anyonic case.

\newsec{Statistics and the Witten Index}

In $2+1$ dimensions, as is well-known, anyon statistics are possible.
The standard definition of statistics
in $2+1$ dimensions amounts to associating a phase of
$\exp(2\pi i j)$ under the interchange of two objects of spin $j$.
Since the rotation group in $2+1$ dimensional theories allows states 
of any real spin, the statistics in $2+1$ dimensions
can yield any complex phase under particle interchange.

Fundamentally, this definition of statistics is made with
reference to the topological properties of the paths of particles
in configuration space; anyon statistics arise 
from the general assignment of 
phases to the topologically distinct sectors of multi-particle
configuration 
space. In the framework of the braid group, anyons
correspond to one-dimensional representations of the 
braid group.\foot{One can also
construct non-abelian statistics based on higher
dimensional representations of the braid group, using, for example,
couplings to non-abelian gauge fields with Chern-Simons terms.  The
methods and results of this paper that show that the index is integral
in ordinary
anyon theories are equally valid in the non-abelian case.  For
simplicity, we will 
use ordinary anyons if a particular example is needed.}

Now in supersymmetric theories with anyons, supersymmetry pairs
finite energy states of spin 
$j$ and $j+{1\over 2}$ \ref\suany{Z. Hlousek and D. Spector, {\it 
Nucl. Phys.},  {\bf B344} (1990) 763.}. Of course, states of zero energy
are 
still supersymmetry singlets. Consequently, the natural definition of 
the Witten index $I_A$
for a supersymmetric theory in $2+1$ dimensions is
\eqn\Ifrac{I_A = \tr \bigl( e^{2\pi i J}e^{-\beta H}\bigr)~,}
where $J$ is the generator of spatial rotations, and thus 
$e^{2\pi i J}$ generalizes $(-1)^F$. Because of the pairing of 
states provided by supersymmetry, $I_A$ is a topological index, with
all the standard topological properties that we associate with the
Witten
index in more familiar contexts. The index $I_A$ only 
receives contributions from zero energy states.
If $I_A$ is non-zero, supersymmetry is not broken.
The index $I_A$ is unchanged by supersymmetric deformations of the
parameters
(provided, as usual, that said deformations do not radically alter the
Hilbert space of the theory --- see below),
and can be reliably and exactly 
calculated in any approximation scheme that respects supersymmetry.
In particular, neither the addition of massive 
particles nor the deformation of the parameters of the theory 
in a way that does not
change the asymptotic behavior of the field theory potential
can alter the value of the index.

Let us consider the application of this index to a
simple model.  
This example does not deal directly with anyonic features,
but serves as a preview of the argument in the next section.
The model is given by the Lagrangian
\eqn\slag{{\cal L} =\int{d^2\theta 
   \bigl({1\over 2}D^\alpha\Phi D_\alpha \Phi 
   + {g\over 3}\Phi^3 - \lambda \Phi \bigr)~,}}
where $\Phi$ is a real superfield, whose physical components are a real 
scalar $\phi$  and a two-component spinor $\psi$.
In terms of the physical component fields, the Lagrangian becomes
\eqn\complag{{\cal L}= {1\over 2}\partial^\mu \phi \partial_\mu \phi
   +{i\over 2}{\bar \psi}\gamma^\mu \partial_\mu \psi
   +\bigl(g\phi^2 - \lambda)^2 + 2g\phi{\bar\psi}\psi~.}
 Without loss of generality, we take
$g>0$. The constant $\lambda$ is real.

This theory has two phases: one in which its $Z_2$ symmetry is manifest,
when
$\lambda<0$; and one in which the $Z_2$ is broken, when $\lambda>0$.
Since the index $I_A$ is unchanged by changes in $\lambda$, we calculate
$I_A$ in the easier phase.  For $\lambda<0$, there are no classical
zero energy configurations, and so the index is zero.  
In the $Z_2$-broken phase, one still has  $I_A=0$
There are now two 
classical minima with zero energy,
\eqn\minima{\phi = \pm \sqrt{\lambda\over g}~.}
The excitations about each of these minima are all massive,
and so $I_A=0$ implies 
\eqn\Isum{e^{2\pi i j_1}+e^{2\pi i j_2} = 0~,}
where $j_1$ and $j_2$ are the spins
of the states associated with the two minima.
Thus these two classical vacua correspond to two quantum
states that have spins that differ by a half-integer.  
(If there were no possibility of anyons in $2+1$ dimensions,
one could conclude on the basis of
index arguments alone that one of these states is fermionic and 
the other bosonic.)

This is as far as the index alone can takes us.  One can gain further
insight into the theory by combining 
index-based results with other pieces
of information about the theory (e.g., requiring that the ground 
state leave rotational invariance unbroken, or invoking the connection
between classical vacua and kinks
\ref\fermions{D. Finkelstein and J. Rubinstein, {\it J.
Math. Phys.} {\bf 9} (1968) 1762\semi
F. Wilczek, {\it Phys. Rev. Lett.} {\bf 49} (1982) 957.}
\ref\fff{D. Spector,  {\it Mod. Phys. Lett.} {\bf A9} (1994)
2245-2251.}).
We do not delve into these possibilities here.

Equating the index in distinct phases is a powerful tool, and we now
use such a method to analyze the index in supersymmetric anyon theories.

\newsec{The Anyonic Index is Integral}

In this section, we will show that the Witten index in anyonic theories
is integral, despite anyon effects.\foot{Whether it might be
non-integral due 
to other effects such as those discussed in \akhoury\ is not at issue
here.}
We will do this by formulating the anyons using a Chern-Simons term,
and then by invoking the topological invariance of the
Witten index to find a regime in which, despite the presence of a
Chern-Simons term, the index can be demonstrated easily to be integral.
This result holds whether
the gauge fields in the Chern-Simons term are auxiliary or dynamical.

Our plan is as follows.  We first consider an apparently simple argument 
that it
is implausible for the index to be non-integral, but discover that it
is too naive, and indeed is unsound and logically inconsistent.
Understanding the flaws
in this argument, however, enables us then to develop
a valid argument that the index remains integral even in the presence of
Cherns-Simons terms.

Consider, therefore, the case that space is given by $R^2$.  How might
the index depend on the Chern-Simons coefficient in this case?
On $R^2$, the (abelian) Chern-Simons coefficient is not quantized, 
and so we may apparently vary it without varying the index.  As this
coefficient
is varied, the spins of the fields, and in turn the statistics of the
particles they create, also vary continuously.  For the index
to remain unchanged when this occurs, it seems plausible that only those
states
whose statistics are unaffected by the change in this coefficient would
make a net contribution to the index (i.e., the gauge neutral states),
while
the states with non-zero gauge charge would cancel each other.  The
gauge
neutral states, in turn, all make integral contributions to the index,
and so
it would be natural to conjecture that the Witten index is integral in 
anyonic theories, as it seems implausible that the index could  be
fractional
and yet remain unchanged as the Chern-Simons coefficient varied.
(Indeed, if the gauge charges of the matter fields in a $U(1)$ gauge
theory
are  commensurate, one can vary the Chern-Simons coefficient to
a point where all the perturbative states have an integer or
half-integer
spin, thereby producing an integer index.)

Unfortunately, this argument, even as a plausibility argument, is
flawed.
Changing the Chern-Simons coefficient on $R^2$
changes the theory discontinuously; infinitesimal changes in the
coefficient
can change whether the wavefunctions live on a finite or infinite cover
of the plane, for example.
Furthermore,
in order to define the theory properly, indeed to calculate quantities
such as 
the Witten index, one typically formulates the theory with space having
the 
topology of a torus or other Riemann surface.  However, on such a
surface, the
Chern-Simons coefficient is quantized, and therefore cannot be
continuously
varied.   (For a detailed treatment of the constraints on the appearance
of exotic statistics on Riemann surfaces, 
see \ref\rsany{T.D. Imbo and J. March-Russell, {\it Phys. Lett. }{\bf
B252}
  (1990) 84.}.)
Thus in order to address our problem, we need
 an approach that does {\it not } depend on varying the value
of the Chern-Simons coefficient, and ideally we would like an approach
that does not depend on the choice of spatial manifold.  As a bonus,
it would be nice if our method applied to
both abelian and non-abelian Chern-Simons terms.
Fortunately, a method that meets all three of these criteria exists.

Our goal, then, is
to demonstrate that, in the presence of a Chern-Simons
term,  the index $I_A$,
even though it generically
receives non-integral contributions, is integral.\foot{That is,
integral despite anyon effects.  A mechanism such as that
of \akhoury\ may
still occur, as in any number of dimensions.  Here, we are only
interested in the particular effects of anyons.}
This does not rule out states of fractional spin and statistics, of
course;
rather, it is that the net contribution of such anyonic states is
nonetheless
integral.

The argument is as follows. Consider the 
index $I_A$ in a theory in which a $U(1)$ or non-abelian Chern-Simons 
term endows the fields with anyon spin and statistics. We may add a
massive 
superfield to this theory without altering the index. We therefore add a 
massive 
field which has a $U(1)$ or non-abelian 
gauge charge and a quartic (or hexadic, since this
is $2+1$ dimensions) scalar potential.

Because the behavior of the potential at large field strength will be 
dominated by the quartic or hexadic terms, changing the coefficient
of the quadratic term will not change the index \windex . Consequently,
we vary this coefficient in the potential until it is 
negative, thus producing the Higgs mechanism. 
In the Higgs phase, anyon behavior is {\it not} induced in the
perturbative spectrum, as observed in \ref\wenzee{X. G. Wen and A. Zee,
{\it J. de Physique} {\bf 50} (1989) 1623.} and discussed in detail
in \suany .  The interested reader should consult these references, but
the essential ideas are easy to summarize here.
In the Higgs phase, there is no Aharonov-Bohm effect
associated with electromagnetic point charges; this is because the 
solutions to the equations of motion for the
combination $B+\mu A^0$ (where $B$ is the magnetic field, $\mu$ the
Chern-Simons mass, and $A^0$ the temporal component of the gauge field)
are now exponentially damped, with a characteristic distance given by
the inverse of the Higgs mass.  In addition, the gauge field propagator
is short range, and so there are no anomalous spin contributions 
generated in perturbation theory.  Consequently the fields of the theory
retain their canonical spin and statistics, and the perturbative
spectrum therefore consists of states of 
integral and half-integral spin. 
Thus, in the Higgs phase, the index $I_A$ is 
necessarily integer-valued. But the index has the same value in the
Higgs 
(broken) phase 
as in the anyon (unbroken) phase \windex, and so the
original theory, although it contains 
anyons, 
must have integral index.

In preparation for the next section, we note that
since the Higgs field has no direct coupling to
the other matter fields, the only effect of the non-zero
Higgs expectation value on the perturbative calculation of the 
index is to modify the gauge multiplet masses and to cause the
charged fields to have the canonical statistics.  Thus the only
difference between the Higgs and anyon phase calculations of the
index is the difference in the statistics.  This shows us, then,
that calculating the index using the naive statistics rather than
the correct anyon statistics will yield the correct value of the
index in the anyon phase.  We return to this point in the next section.

These results represent a significant constraint on the behavior
of supersymmetric theories in $2+1$ dimensions.  To conclude our 
discussion of these results, it is worth making 
two additional observations.

First, it is known that in the Higgs phase of 
abelian Chern-Simons models, there may be non-perturbative states, such
as 
vortices, with anyon spin. Since perturbative index calculations are
exact, the net contribution of such non-perturbative states to the index
must be zero, and so such states cannot change our conclusion.
Indeed, the vortices are massive, and so obviously 
give zero contribution to the index, as expected.

Second, note that the
above argument that $I_A$ is integral clearly applies as well to
the case that the exotic particle statistics arise from
a non-abelian Chern-Simons term.  Thus our finding is not restricted
to the simple case of anyons, but encompasses their non-abelian
generalization as well. 

\newsec{An Alternative Index}

The results of the preceding section, which showed that the
index is integral despite non-integral contributions in the sum, suggest
the possibility of defining an alternative index
that is automatically an integer
but that has the same value as the Witten index $I_A$
defined above.  To this end, we introduce a notion of statistics
for quantized field theories based on how fields are quantized.
Let us define an operator $\Omega$ with the properties
that $\Omega^2 =1$ and that 
$\Omega$ (anti)commutes with those fields that are quantized with
canonical 
(anti)commutation relations.  
In addition, we specify that the vacuum have eigenvalue $+1$ under the
action of $\Omega$; without this, we have only specified the relative
value of $\Omega$ on the states in the Hilbert space.
We use this operator to give an
alternative definition of statistics in $2+1$
dimensional quantized field theory.  The operator $\Omega$
has eigenvalues $+1$ and $-1$, and
thus provides a twofold grading of the fields, 
of the creation and
annihilation operators associated with the fields, and hence of the Fock
space as well.
Since one can formulate a supersymmetric theory by quantizing
superfields
in superspace, and since a superfield can be expanded in 
terms of a Grassman parameter $\theta$, each superfield, obviously,
pairs
canonically commuting and anticommuting fields, as therefore does the
supercharge.
Therefore $\Omega$ anticommutes with the supersymmetry charge,
and
supersymmetry pairs Fock space states of $\Omega = +1$
with Fock space states of $\Omega = -1$.

Since supersymmetry pairs states of $\Omega = +1$ with states
of $\Omega = -1$,
we are led to define the alternative index
\eqn\Iw{I_Z = \tr \bigl( e^{-\beta H} \Omega \bigr)~.}
Since $\Omega^2=1$, the index $I_Z$ is manifestly integral.
Since $\Omega$ anticommutes with the supersymmetry
generators, the index $I_Z$ also possesses
all the expected properties of a Witten index:
it can receive net non-zero contribution only from the
zero energy states; when it is 
non-zero, supersymmetry is unbroken; it is unchanged by supersymmetric 
deformations of the parameters of a theory; and it is reliably and
exactly 
calculated in supersymmetric approximation schemes.
Note also that $I_Z$ is the naive index that would be calculated
by someone analyzing a field theory with a Chern-Simons gauge field
mass term who did not know that the Chern-Simons term generates anyon
statistics for charged fields. 

In higher dimensions, the two indices $I_A$ and $I_Z$ are manifestly
identical to
each other, term by term in the sums, due to the spin-statistics
theorem.
Here, however, the indices  $I_A$ and $I_Z$ are 
in principle qualitatively distinct from each other.
One way of understanding the difference between these two indices
is that $I_Z$ merely keeps track of the pairing of states into
superpartnerships, whereas $I_A$ also keeps track of the spin associated
with
each superpartner pairing.  Thus it is that $I_Z$ is a sum of integers,
while $I_A$ is not, which was the reason for defining $I_Z$ in the first
place.  
It is important, however, to recognize that the 
quantity $I_Z$ is defined in a less general context than  $I_A$ is,
since $I_Z$ makes explicit reference to the fields of a theory and
to the quantization prescription, whereas $I_A$ only 
makes reference to the Hilbert space of states and two of the operators
($J$ 
and $H$) that act on that space.

Thus, by introducing an alternate definition of statistics,
we have now identified two 
distinct Witten indices for $2+1$ dimensional theories. 
Nonetheless, these two indices are always equal in value.
Recall our calculation of the index $I_A$ in the previous section, in
which we included a Higgs field which manifestly did not alter the
index in the unbroken phase, and which enabled us to relate the index
in two different phases of the theory.  
This same technique is the basis for establishing
the equality of $I_Z$ and $I_A$.
Since in any Chern-Simons theory, the naive 
statistics are equivalent to the statistics defined 
via the operator $\Omega$ introduced above,
in the Higgs phase, where anyon spin and statistics are not
generated for the fields of a theory, the two  
indices $I_A$ and $I_Z$ are automatically equal to each other.
In addition, $I_Z$, like $I_A$, is 
unchanged by changes in the Higgs field mass
parameter, and thus each index independently has the same 
value in the Higgs phase as it does in the anyon phase. Therefore, 
we see that $I_A$ and $I_Z$ are equal in the anyonic phase of the theory 
with the added Higgs superfield, and consequently we see that $I_A =
I_Z$
in the original theory, too, since the added
Higgs superfield does not change
the value of either index.  And since $I_Z$ is automatically an integer,
we see again by this approach that
$I_A$ must be integral, too.

\newsec{Concluding Thoughts: Review, Discussion, and Speculation}

Using the equivalence of the index in Higgs and anyon phases, we have
demonstrated that the Witten index is always integral in supersymmetric
anyon theories.  The arguments apply to theories coupled to abelian
and non-abelian Chern-Simons terms.  We see, too, that an alternative
index defined in terms of commutation properties of the fields, while
less fundamental, does naturally produce this integer value.
The equality of $I_A$ and $I_Z$ can be a useful calculational tool;
since $\Omega$ eigenvalues, unlike $J$ eigenvalues, cannot vary
continuously even in $2+1$ dimensions, the index $I_Z$ is easier to 
calculate reliably.

It would be most intriguing to find an example where an integral index
$I_A$ arises from a truly exotic combination of vacuum statistics,
not simply from pairwise cancellations, although
the possibility of this is severely constrained by the equality
of $I_A$ and $I_Z$.
In addition,
combining index calculations with an analysis of the spectrum
based on other techniques might yield interesting relationships
among the states in a theory with anyons in a relatively easy
and efficient manner, potentially even yielding an alternative
avenue for identifying the appearance of anyon behavior itself.
Even more speculatively, perhaps one could find a condensed
matter system which is effectively planar and exhibits a
phenomenological supersymmetry to which our Witten index
results could be applied.

\bigbreak\bigskip\bigskip\centerline{{\bf Acknowledgments}}\nobreak
I thank Al Shapere for discussions. 
I also thank the Theory Group at NIKHEF for its hospitality.
This work was
supported in part by NSF Grant. No. PHY-9509991.
             
\listrefs
\bye